\documentclass[aps,pre,twocolumn,superscriptaddress,showpacs]{revtex4-1}
\usepackage[dvips]{graphicx}
\usepackage{amsmath}

\usepackage[addmarkup=uline]{changes}
\definechangesauthor[color=blue]{SY}

\begin{document}

\title{Time and Observables in Covariant Quantum Theory}

\author{Natalia Gorobey}
\affiliation{Peter the Great Saint Petersburg Polytechnic University, Polytekhnicheskaya
29, 195251, St. Petersburg, Russia}

\author{Alexander Lukyanenko}\email{alex.lukyan@mail.ru}
\affiliation{Peter the Great Saint Petersburg Polytechnic University, Polytekhnicheskaya
29, 195251, St. Petersburg, Russia}

\begin{abstract}
A modification of the covariant theory is proposed in which the self-energy
of the system, corresponding to time-like degrees of freedom in the
configuration space, preserves the classical law of change in quantum
theory. As a result, proper time in covariant quantum theory takes on a
dynamic meaning. As applications of the new formalism, a modification of the
relativistic quantum mechanics of a scalar particle and a homogeneous model
of the 
%Friedmann 
universe is considered.
\end{abstract}

\maketitle

%\begin{document}
%
%\title{Time and Observables in Covariant Quantum Theory}
%\author{Natalia Gorobey, Alexander Lukyanenko}
%\email{alex.lukyan@mail.ru}
%\affiliation{Peter the Great Saint Petersburg Polytechnic University, Polytekhnicheskaya
%29, 195251, St. Petersburg, Russia}
%
%\begin{abstract}
%A modification of the covariant theory is proposed in which the self-energy
%of the system, corresponding to time-like degrees of freedom in the
%configuration space, preserves the classical law of change in quantum
%theory. As a result, proper time in covariant quantum theory takes on a
%dynamic meaning. As applications of the new formalism, a modification of the
%relativistic quantum mechanics of a scalar particle and a homogeneous model
%of the Friedmann universe is considered.
%\end{abstract}
%
%\date{\today }
%
%%\pacs{}
%
%%\begin{multicols}{2}
%%\narrowtext
%
%%%%%%%%%%%%%%%%%%%%%%%%%%%%%%%%%%%%%%%%%%%%%%%%%%%%%%%%%%%%%%%%%%%%%%%%

%%\bigskip

%\section{ABSTRACT}

\section{INTRODUCTION}

In theories based on the principle of covariance, including
reparametrization of time, there is a problem with the determination of this
time after quantization. Here we focus on the covariant approach to
quantization in which the Batalin-Fradkin-Vilkovysky
theorem (BFV) \cite{1,2,3} and the concept of BRST invariance
\cite{4,5} play the central role. However, for a simplest covariant system --- a
relativistic particle, the result of this formalism reduces to the
well-known representation of the Feynman propagator in the form of an
integral over the particle's proper time \cite{6}. This representation of
the Green's function for the Dirac and Klein-Gordon operators was first
introduced by Fock \cite{7} and Schwinger \cite{8} and is now widely used
for calculations \cite{9}. The Feynman propagator serves to solve the
scattering problem in relativistic quantum mechanics in the framework of
perturbation theory \cite{10}. It should be expected that in the general
case the propagator contains a multiple integral over the parameters of the
system's proper time (multi-arrow time in the theory of gravity \cite{11}),
as well as over the remaining parameters of the full symmetry group. In this
paper, a simplified version of the construction of a propagator of covariant
theory is proposed: the amplitude of the transition is determined on the
space of the symmetry group\textbf and then it is integrated over all group
parameters. Thus, proper time is also excluded from the dynamics. To restore
the dynamic meaning of proper time, based on this integral representation of
the covariant propagator, a modification of the covariant quantum theory is
proposed in this paper. The proposed modification can be considered as an
approximation in which certain dynamic variables (in each case it is clear
which ones) are limited by the classical laws of their motion. Its result is
precisely the removal of integration in proper time. Formally, the
approximation is reduced to the linearization of quadratic in momenta
(Hamiltonian) constraints with respect to the corresponding canonical
momenta and the appearance of $\delta $-functions that remove integration
over proper time. Together with the time parameter in a modified theory, an
observable is associated with it  that has the physical meaning of the
system's own energy, corresponding to time-like degrees of freedom in the
configuration space. As examples, the quantum mechanics of a scalar
relativistic particle and a homogeneous model of the Friedman universe are
considered.

In the section \ref{parameters}, a simplified version of the covariant quantum theory is
formulated, which then will be the basis for its modification. In the section \ref{propagator},
we consider a modification of relativistic quantum mechanics, in
which the $x_{0}$ -coordinate of a particle in Minkowski space serves as the
evolution parameter, and the corresponding observable is the particle's own
energy. In the section \ref{modified propagator}, we consider a modified quantum theory of the
homogeneous Friedmann universe, in which the radius of the universe serves
as the evolution parameter, and the energy of space is the observable
associated with it. In this case, the pair of variables (the radius of the
universe and the energy of the space) can exchange their roles (time and the observable).
This result is consistent with the proposal to consider space energy density
as a parameter of time in cosmology \cite{12}.

\section{PARAMETERS OF THE SYMMETRY GROUP AS DYNAMIC VARIABLES}
\label{parameters}

We take the initial action of the covariant theory in the canonical form \cite%
{13}
\begin{equation}
I=\int_{0}^{1}d\tau \left[ p_{k}\overset{\cdot }{q}_{k}-\lambda _{a}\varphi
_{a}\right] ,  \label{1}
\end{equation}%
where the Hamiltonian $h\left( p,q\right) $ is set equal to zero. Moreover,
in the set of constraints $\varphi _{a}\left( p,q\right) $ we assume the
presence of quadratic in momenta (Hamiltonian) constraints, as is the case
of a relativistic particle and cosmology that interest us. The constraints
form a closed Lie algebra with respect to the Poisson brackets,
\begin{equation}
\left\{ \varphi _{a},\varphi _{b}\right\} =C_{abd}\varphi _{d},  \label{2}
\end{equation}%
where the structural "constants" $C_{abd}$ generally depend on the canonical
variables. In the theory of gravity, they depend on the $3D$ metric \cite%
{14}. Here, for simplicity, we consider them to be constants. They obey
Jacobi's identity:
\begin{equation}
C_{bcd}C_{adp}-C_{abd}C_{dcp}-C_{acd}C_{bdp}=0.  \label{3}
\end{equation}%
The action Eq. (\ref{1}) is invariant with respect to gauge transformations $\left(
\varepsilon _{d}\left( 0\right) =\varepsilon _{d}\left( 1\right) =0\right) $,
\begin{equation}
\delta q_{k}=\varepsilon _{a}\left\{ q_{k},\varphi _{a}\right\} ,\delta
p_{k}=\varepsilon _{a}\left\{ p_{k},\varphi _{a}\right\},  \label{4}
\end{equation}%
provided that the Lagrange multipliers are converted as follows:
\begin{equation}
\delta \lambda _{a}=\overset{\cdot }{\varepsilon }_{a}-C_{bda}\lambda
_{b}\varepsilon _{d}.  \label{5}
\end{equation}%
We introduce the explicit dependence of the Lagrange multipliers $\lambda
_{a}$ on the group parameters $s_{a}$ for which $\varepsilon _{a}=\delta
s_{a}$ are infinitesimal variations, and write the relation Eq. (\ref{5}) in the form
of a functional differential equation
\begin{equation}
\frac{\delta \lambda _{a}\left( \tau \right) }{\delta s_{d}\left( \tau
^{\prime }\right) }=\delta _{ad}\frac{d}{d\tau }\delta \left( \tau -\tau
^{\prime }\right) -C_{bda}\lambda _{b}\delta \left( \tau -\tau ^{\prime
}\right) ,  \label{6}
\end{equation}%
which is supplemented by the initial condition $\lambda _{a}\left( 0\right)
=0$. The solution of Eq. (\ref{6}) is linear with respect to the
velocities $\overset{\cdot }{s}_{a}$:
\begin{equation}
\lambda _{b}=\overset{\cdot }{s}_{a}\Lambda _{ab},  \label{7}
\end{equation}%
where
%, to the second order, in the expansion in a Taylor series,
the Teylor series of $\Lambda _{ab}$ in $s$ up to the second order is
\begin{equation}
\Lambda _{ab}=\delta _{ab}-\frac{1}{2!}C_{adb}s_{d}+\frac{1}{3!}%
C_{ad^{\prime }b^{\prime }}C_{b^{\prime }db}s_{d^{\prime }}s_{d}+....
\label{8}
\end{equation}%
Thus, the original coordinate set $q_{k}$ is supplemented with group
parameters $s_{a}$ that are now explicitly present in the action:
\begin{equation}
I=\int_{0}^{1}d\tau \left[ p_{k}\overset{\cdot }{q}_{k}-\overset{\cdot }{s}%
_{b}\Lambda _{ba}\left( s\right) \varphi _{a}\right] .  \label{9}
\end{equation}%
It remains to write down the action Eq. (\ref{9}) in canonical form:
\begin{equation}
I=\int_{0}^{1}d\tau \left[ p_{k}\overset{\cdot }{q}_{k}+P_{a}\overset{\cdot }%
{s}_{a}-\eta _{a}\Phi _{a}\right] ,  \label{10}
\end{equation}%
where
\begin{equation}
\Phi _{a}=P_{a}+\Lambda _{ab}\varphi _{b}.  \label{11}
\end{equation}%
A set of constraints obeys the simple algebra of Poisson brackets
\begin{equation}
\left\{ \Phi _{a},\Phi _{b}\right\} =0  \label{12}
\end{equation}%
at the condition
\begin{equation}
\frac{\partial \Lambda _{ap}}{\partial s_{q}}-\frac{\partial \Lambda _{sp}}{%
\partial s_{a}}+\Lambda _{ab}\Lambda _{qr}C_{brp}=0.  \label{13}
\end{equation}%
The relation (\ref{13}) holds in our case with the indicated accuracy as a
consequence of the Jacobi identity Eq. (\ref{3}).

We quantize the theory based on the new canonical form of the action Eq. (\ref{10}).
For the wave function $\psi \left( s,q\right) $, we obtain the system of
wave equations
\begin{equation}
i\hbar \frac{\partial \psi }{\partial s_{a}}=\Lambda _{ab}\left( s\right)
\varphi _{b}\left( q,\widehat{p}\right) \psi ,  \label{14}
\end{equation}%
which have the form of an equation of evolution on the space of a symmetry
group. In our formulation, the covariance condition is that, since the group
parameters $s_{a}$ are unobservable, the solution of the system Eq. (\ref{14}) should
be integrated over the entire range of their variation. So, in the presence
of an abelian gauge subgroup, the corresponding part of the equations of the
system Eq. (\ref{14}) has the form
\begin{equation}
\frac{\partial \psi }{\partial s_{a}}=\frac{\partial \psi }{\partial q_{a}},
\label{15}
\end{equation}%
where $q_{a}$ are (longitudinal) coordinates, on which the wave function
should not depend. According to Eq. (\ref{15}), the wave function depends on the
sum $q_{a}+s_{a}$ so that integration over $s_{a}$ in the range from $%
-\infty $ to $+\infty $ excludes the dependence on $q_{a}$.

The group parameters corresponding to the Hamiltonian constraints have the
meaning of the proper times of the system. One should integrate over them
within the limits $\left[ 0,+\infty \right) $ \cite{6}. It is these
integrals that are the subject of concern in our subsequent constructions,
and we will still refine this construction in the process of its application.

\section{MODIFIED PROPAGATOR FOR RELATIVISTIC PARTICLE}
\label{propagator}

We take the initial action of a relativistic particle in a parameterized
form (speed of light $c=1$) in a form
\begin{equation}
\widetilde{I}=\int_{0}^{1}d\tau \left[ \frac{1}{4}\left( \frac{\overset{%
\cdot }{x}_{0}^{2}}{N}-\frac{\overset{\cdot }{x}_{i}^{2}}{N}\right)
+m^{2}N+e\varphi \left( x_{0},x_{i}\right) \overset{\cdot }{x}_{0}\right] ,
\label{16}
\end{equation}%
where the dot denotes the derivative with respect to the parameter $\tau $
on its world line. Here we restrict ourselves to the interaction with the
electric field of strength $E_{i}=-\partial _{i}\varphi $. The Hamiltonian
constraint
\begin{equation}
H=\left( p_{0}-e\varphi \right) ^{2}-p_{i}^{2}-m^{2}  \label{17}
\end{equation}%
is quadratic in momenta. Our task in the future is to remove the integration
over the particle's proper time
\begin{equation}
s=\int_{0}^{1}d\pi N\left( \tau \right)   \label{18}
\end{equation}%
in the Fock-Shwinger (FS) representation for the Feynman propagator in an
external electric field. We modify the dynamics of the coordinate $x_{0}$ of
Minkowski space, adding to the action (\ref{16}) its variation generated by
the infinitesimal time shift of $x_{0}$,
\begin{equation}
\delta x_{0}=-\frac{\overset{\cdot }{x}_{0}}{N}\varepsilon .  \label{19}
\end{equation}%
A new action,
\begin{eqnarray}
&&
\int_{0}^{1}d\tau \left[ \frac{1}{4}\frac{\overset{\cdot }{x}_{0}^{2}}{N}%
\left( 1-\frac{\overset{\cdot }{\varepsilon }}{N}\right) -\frac{1}{4}\frac{%
\overset{\cdot }{x}_{i}^{2}}{N}+m^{2}N\right.  \notag \\
&&\left. +e\varphi \overset{\cdot }{x}_{0}+e\partial _{i}\varphi \frac{%
\overset{\cdot }{x}_{0}\overset{\cdot }{x}_{i}}{N}\varepsilon \right],
\label{20}
\end{eqnarray}%
contains the time shift $\varepsilon $ as an independent variable. To go
to the canonical form of the modified theory, we find the canonical
momenta,
\begin{equation}
p_{0}=\frac{1}{2}\frac{\overset{\cdot }{x}_{0}}{N}\left( 1-\frac{\overset{%
\cdot }{\varepsilon }}{N}\right) +e\varphi +e\partial _{i}\varphi \frac{%
\overset{\cdot }{x}_{i}}{N}\varepsilon ,  \label{21}
\end{equation}
\begin{equation}
p_{i}=-\frac{1}{2}\frac{\overset{\cdot }{x}_{i}}{N}+e\partial _{i}\varphi
\frac{\overset{\cdot }{x}_{0}}{N}\varepsilon ,  \label{22}
\end{equation}
\begin{equation}
-P_{\varepsilon }=-\frac{1}{4}\frac{\overset{\cdot }{x}_{0}^{2}}{N^{2}},
\label{23}
\end{equation}%
and a modified Hamiltonian,
\begin{equation}
\widetilde{h}{=}N\left[ 2\sqrt{P_{\varepsilon }}\left( p_{0}{-}e\varphi \right)
{-}\left( p_{i}+2e\partial _{i}\varphi \varepsilon \sqrt{P_{\varepsilon }}%
\right) ^{2}{-}P_{\varepsilon }{-}m^{2}\right].  \label{24}
\end{equation}%
%Now
The Hamiltonian function is linear in the momentum $p_{0}$.
%, and
The pair of canonical variables $\left( \varepsilon ,P_{\varepsilon }\right)$
%has appeared, in which
appears and  $\varepsilon $ will be considered as the canonical momentum.
The canonical coordinate $P_{\varepsilon }$ has the physical meaning of the
particle's own energy.
%We will replace
After the replacement
\begin{equation}
\widetilde{p}_{i}=p_{i}+2e\partial _{i}\varphi \varepsilon \sqrt{%
P_{\varepsilon }},  \label{25}
\end{equation}%
%After that,
the canonical form of the modified action takes the form
\begin{equation}
\widetilde{I}=\int_{0}^{s}ds\left[ \widetilde{p}_{i}\overset{\cdot }{x}%
_{i}+p_{0}\overset{\cdot }{x}_{0}-\varepsilon \left( \overset{\cdot }{P}%
_{\varepsilon }+2e\partial _{i}\varphi \overset{\cdot }{x}_{0}\sqrt{%
P_{\varepsilon }}\right) -\widetilde{H}\right] ,  \label{26}
\end{equation}%
where,
\begin{equation}
\widetilde{H}=2\sqrt{P_{\varepsilon }}\left( p_{0}-e\varphi \right) -%
\widetilde{p}_{i}^{2}-P_{\varepsilon }-m^{2}.  \label{27}
\end{equation}%
Here we proceed to the parametrization of the world line of the particle with its
proper time Eq. (\ref{18}). The following dynamic constraint is also added to the action Eq. (\ref{26}):
\begin{equation}
\overset{\cdot }{P}_{\varepsilon }+2e\partial _{i}\varphi \overset{\cdot }{x}%
_{i}\sqrt{P_{\varepsilon }}=0,  \label{28}
\end{equation}%
which represents the classical law of particle energy change in an external
electric field, and the momentum $\varepsilon $ plays the role of the
Lagrange multiplier.

In the formal representation of the modified propagator for a particle as a
functional integral on the phase space,
\begin{equation}
\widetilde{K}=\int_{0}^{\infty }ds\int \prod\limits_{s,i}\frac{dp_{0}dx_{0}}{%
2\pi \hbar }\frac{d\widetilde{p}idx_{i}}{2\pi \hbar }\frac{dP_{\varepsilon
}d\varepsilon }{2\pi \hbar }\exp \left\{ \frac{i}{\hbar }\widetilde{I}%
\right\} ,  \label{29}
\end{equation}%
after functional integration over momenta, we obtain the product of $\delta $%
-functions in the integral on the augmented configuration space:
\begin{eqnarray}
\widetilde{K} &=&\int_{0}^{\infty }ds\int \prod\limits_{s,i}\frac{dx_{i}}{%
\sqrt{2\pi \hbar }}\frac{dx_{0}}{\sqrt{2\pi \hbar }}\frac{dP_{\varepsilon }}{%
\sqrt{2\pi \hbar }}\delta \left( \overset{\cdot }{x}_{0}-2\sqrt{%
P_{\varepsilon }}\right)  \notag \\
&&\times \delta \left( \overset{\cdot }{P}_{\varepsilon }+2e\partial _{i}\varphi
\overset{\cdot }{x}_{i}\sqrt{P_{\varepsilon }}\right)  \notag \\
&&\times \exp \left\{ \frac{i}{\hbar }\int_{0}^{s}ds\left[ \overset{\cdot }{x}%
_{i}^{2}+P_{\varepsilon }+m^{2}+e\varphi \right] \right\} .  \label{30}
\end{eqnarray}%
The propagator Eq. (\ref{30}) serves as the basis for a new dynamic interpretation of
relativistic quantum mechanics. For definiteness, we consider the scattering
problem. In the absence of an external electric field, the integrals over
all internal values of the coordinates $x_{0}$ and $P_{\varepsilon }$ are
removed and the product of the $\delta $-function remains
\begin{equation}
\delta \left( \frac{x_{0}^{\prime }-x_{0}}{s}-2\sqrt{P_{\varepsilon }}%
\right) \delta \left( \frac{P_{\varepsilon }^{\prime }-P_{\varepsilon }}{s}%
\right) .  \label{31}
\end{equation}%
Then, the integral over the proper time in Eq. (\ref{30}) is removed by the first
$\delta$-function. Thus, the physical parameter of time, as expected, is $x_{0}$. In this
asymptotic region, according to the Hamiltonian constraint Eq. (\ref{27}), the
wave equation has the form of the Schr\"{o}dinger equation
\begin{equation}
i\hbar 2\sqrt{P_{\varepsilon }}\frac{\partial \psi }{\partial x_{0}}=\left[
-\hbar ^{2}\Delta +P_{\varepsilon }+m^{2}\right] \psi ,  \label{32}
\end{equation}%
in which the parameter $x_{0}$ plays the role of the classical time
parameter. However,  if we put $P_{\varepsilon
}=E$, it gives the relativistic dispersion law $E^{2}-p_{i}^{2}-m^{2}=0$ for a de Broglie plane wave. In
the interaction region, the integrals over all internal values of the
coordinates $x_{0}$ and $P_{\varepsilon }$ are also removed, but now we
obtain the proper time $s$ in the form of the functional of the trajectory $%
x_{i}\left( s\right) $ of the particle. Therefore, the integral over the
proper time in Eq. (\ref{30}) should be located to the right of the functional
integral along the particle trajectory. We note that the classical law of
the change in the self-energy of the particle $P_{\varepsilon }$ with time
excludes the processes of annihilation and production of
particle-antiparticle pairs in the approximation considered here.
%\textbf{This is precisely what we do not need in the quantum dynamics of the universe. ????}
A  consistent consideration of these processes is possible only after secondary quantization. In cosmology we want to avoid this.

\section{MODIFIED PROPAGATOR FOR FRIEDMAN'S HOMOGENEOUS UNIVERSE}
\label{modified propagator}

The dynamics of the Friedman universe is determined by the action
\begin{equation}
I=\int_{0}^{1}d\tau \left[ -\frac{a}{2g}\left( \frac{\overset{\cdot }{a}^{2}%
}{N}-N\right) +2\pi ^{2}a^{3}\frac{1}{2}\left( \frac{\overset{\cdot }{\phi }%
_{k}^{2}}{N}-V\left( \phi \right) \right) \right] ,  \label{33}
\end{equation}%
where $g=2G/3\pi $, $G$ is the Newtonian gravitational constant and $a$ is the
radius of the universe. We consider the set of scalar fields $\phi
_{k}$ as matter and $N$ is the ADM lapse function \cite{14}. As in the case of a
relativistic particle, we have here one Hamiltonian constraint and
an additional integration over the proper time in the path integral
representing the wave function of the universe \cite{15}. In the Euclidean
form of quantum theory, this integral diverges due to the uncertainty of the
sign of the action. This theory only makes sense in the semiclassical
approximation. We will further search for the dynamic meaning of this
quantum theory in its modification. An infinitesimal time shift,
\begin{equation}
\delta a=-\frac{\overset{\cdot }{a}}{N}\varepsilon ,  \label{34}
\end{equation}%
leads to a modified action
\begin{eqnarray}
\widetilde{I} &{=}&\int_{0}^{1}d\tau \left[ -\frac{1}{2g}\frac{a\overset{\cdot
}{a}^{2}}{N}\left( 1-\frac{\overset{\cdot }{\varepsilon }}{N}\right) +\frac{a%
}{2g}\left( N+\overset{\cdot }{\varepsilon }\right) \right.  \notag \\
&&\left. {+}\pi ^{2}\left( a^{3}-3a^{2}\overset{\cdot }{a}\varepsilon \right)
\left( \frac{\overset{\cdot }{\phi }_{k}^{2}}{N}-V\left( \phi \right)
N\right) \right].  \label{35}
\end{eqnarray}%
Compared to a relativistic particle, the interaction of a time-like dynamic
variable $a$ with matter is more complex dynamics. For the energy of the
space $P_{\varepsilon }$ from Eq. (\ref{35}) we obtain the equation of motion
\begin{equation}
\overset{\cdot }{P}_{\varepsilon }+3\pi ^{2}a^{2}\overset{\cdot }{a}\left(
\frac{\overset{\cdot }{\phi }_{k}^{2}}{N^{2}}-V\left( \phi \right) \right)
=0.  \label{36}
\end{equation}%
We turn to the canonical form of the modified action. We find the canonical
momenta:
\begin{equation}
p_{a}=-\frac{1}{g}\frac{a\overset{\cdot }{a}}{N}\left( 1-\frac{\overset{%
\cdot }{\varepsilon }}{N}\right) -3\pi ^{2}a^{2}\varepsilon \left( \frac{%
\overset{\cdot }{\phi }_{k}^{2}}{N^{2}}-V\left( \phi \right) \right) ,
\label{37}
\end{equation}
\begin{equation}
p_{\phi _{k}}=2\pi ^{2}\left( a^{3}-3a^{2}\overset{\cdot }{a}\varepsilon
\right) \frac{\overset{\cdot }{\phi }}{N},  \label{38}
\end{equation}
\begin{equation}
P_{\varepsilon }=\frac{a}{2g}\left( \frac{\overset{\cdot }{a}^{2}}{N^{2}}%
+1\right) .  \label{39}
\end{equation}%
The modified Hamiltonian constraint is now linear in momentum $p_{a}$:
\begin{eqnarray}
\!\!\!\!\!\!\!\widetilde{H} &{=}&p_{a}\sqrt{\frac{2gP_{\varepsilon }}{a}-1}+\frac{a}{2g}%
\left( \sqrt{\frac{2gP_{\varepsilon }}{a}-1}-1\right)  \notag \\
&&{+}\frac{1}{2}\left[ \frac{p_{\phi _{k}}^{2}}{2\pi ^{2}\left(
a^{3}-3a^{2}\zeta \right) }+2\pi ^{2}\left( a^{3}{-}3a^{2}\zeta \right)
V\left( \phi \right) \right],  \label{40}
\end{eqnarray}%
where
%indicated:
%
\begin{equation}
\zeta =\overset{\cdot }{a}\varepsilon =\varepsilon \sqrt{\frac{%
2gP_{\varepsilon }}{a}-1}.  \label{41}
\end{equation}%
Here the difficulty arises: the infinitesimal time shift $\varepsilon $,
which serves as the canonical momentum, fells into the denominator. It
will become a problem after quantization. We write, however, a modified
propagator in the form of a formal continual integral in the canonical form,
\begin{eqnarray}
\widetilde{K} &=&\int_{0}^{\infty }ds\int \prod\limits_{s}\frac{dp_{a}da}{%
2\pi \hbar }\frac{d\varepsilon dP_{\varepsilon }}{2\pi \hbar }%
\prod\limits_{k=1}^{\alpha }\frac{dp_{\phi _{k}}d\phi _{k}}{2\pi \hbar }%
\frac{d\lambda d\zeta }{2\pi \hbar }J  \notag \\
&&\times \exp \left( \frac{i}{\hbar }\widetilde{I}\right) ,  \label{42}
\end{eqnarray}%
where
\begin{equation}
\widetilde{I}=\int_{0}^{s}ds\left[ p_{\phi _{k}}\overset{\cdot }{\phi }%
_{k}+p_{a}\overset{\cdot }{a}-\varepsilon \overset{\cdot }{P}_{\varepsilon
}+\lambda \left( \zeta -\overset{\cdot }{a}\varepsilon \right) -\widetilde{H}%
\right] .  \label{43}
\end{equation}%
Here, we took into account the relation Eq. (\ref{41}), $\lambda $ is the
corresponding Lagrange multiplier, and $J$ will be the \textquotedblleft
price\textquotedblright\ for the mentioned complication in the Hamiltonian
constraint. Momentum integration gives:
\begin{eqnarray}
\widetilde{K} &=&\int_{0}^{\infty }ds\int \prod\limits_{s}\frac{%
dP_{\varepsilon }da}{2\pi \hbar }\frac{dp_{a}da}{2\pi \hbar }  \notag \\
&&\times \prod\limits_{k}\frac{d\phi _{k}}{\left( 2\pi \hbar \right)
^{\alpha /2}\left[ 2\pi ^{2}\left( a^{3}-3a^{2}\zeta \right) \right]
^{\alpha /2}}  \notag \\
&&\times \frac{d\lambda d\zeta }{2\pi \hbar }J\delta \left( \overset{\cdot }{%
a}-\sqrt{\frac{2gP_{\varepsilon }}{a}-1}\right) \delta \left( \overset{\cdot
}{P}_{\varepsilon }+\lambda \overset{\cdot }{a}\right)  \notag \\
&&\times \exp \left( -\frac{i}{\hbar }\int_{0}^{s}ds\lambda \zeta \right)
\notag \\
&&\times \exp \left\{ -\frac{i}{\hbar }\int_{0}^{s}ds\left[ \frac{a}{2g}%
\left( \sqrt{\frac{2gP_{\varepsilon }}{a}-1}-1\right) \right] \right\}
\notag \\
&&\times \exp \left\{ -\frac{i}{\hbar }\int_{0}^{s}ds\left[ \pi ^{2}\left(
a^{3}-3a^{2}\zeta \right) \right. \right.  \notag \\
&&\left. \left. \times \left( \overset{\cdot }{\phi }_{k}^{2}-V\left( \phi
\right) \right) \right] \right\} .  \label{44}
\end{eqnarray}%
Thus, in order to get a reasonable expected result, you should put
\begin{equation}
J=\left[ 2\pi ^{2}\left( a^{3}-3a^{2}\zeta \right) \right] ^{\alpha /2}.
\label{45}
\end{equation}%
The integral over $\zeta $ is now taken and gives the $\delta $-function
\begin{equation}
\delta \left( \lambda -3a^{2}\left( \overset{\cdot }{\phi }_{k}^{2}-V\left(
\phi \right) \right) \right) .  \label{46}
\end{equation}%
Then the integral over $\lambda $ is taken. Finally, we obtain a modified
propagator containing two $\delta $-functions:
\begin{eqnarray}
\widetilde{K} &=&\int_{0}^{\infty }ds\int \prod\limits_{s}\frac{%
dP_{\varepsilon }da}{2\pi \hbar }\prod\limits_{k}\frac{d\phi _{k}}{\left(
2\pi \hbar \right) ^{\alpha /2}}\delta \left( \overset{\cdot }{a}-\sqrt{%
\frac{2gP_{\varepsilon }}{a}-1}\right)  \notag \\
&&\times \delta \left( \overset{\cdot }{P}_{\varepsilon }+3a^{2}\overset{%
\cdot }{a}\left( \overset{\cdot }{\phi }_{k}^{2}-V\left( \phi \right)
\right) \right)  \notag \\
&&\times \exp \left\{ -\frac{i}{\hbar }\int_{0}^{s}ds\left[ \frac{a}{2g}%
\left( \sqrt{\frac{2gP_{\varepsilon }}{a}-1}-1\right) \right] \right\}
\notag \\
&&\times \exp \left\{ -\frac{i}{\hbar }\int_{0}^{s}ds\left[ \pi
^{2}a^{3}\left( \overset{\cdot }{\phi }_{k}^{2}-V\left( \phi \right) \right) %
\right] \right\} .  \label{47}
\end{eqnarray}%
These $\delta $-functions remove the integration over $a,P_{\varepsilon }$
in the functional integral. The remaining two $\delta $-functions, that
depend on the boundary values of the variables $a,P_{\varepsilon }$ in the
propagator (any of them), take the integral in proper time and determine the
physical time.

There are two possibilities for this. The most familiar is the choice of the
radius $a$ as a parameter of time with the beginning in the singularity $a=0$%
. Now we need to set the initial state of the universe near the singularity,
as well as the initial value of the energy of the space $P_{\varepsilon }$.
The difficulty is that the Hamiltonian constraint (\ref{40}) and the
corresponding Schr\~{o}dinger equation in quantum theory are connected by
the parameter $\zeta $ to the dynamic constraint Eq. (\ref{39}). To unlock this
connection, we will make further approximations. Near the singularity,
the constraint Eq. (\ref{40}) is simplified:
\begin{equation}
\widetilde{H}\cong \sqrt{\frac{2gP_{\varepsilon }}{a}}p_{a}-\frac{p_{\phi
_{k}}^{2}}{12\pi ^{2}a^{2}\zeta }.  \label{48}
\end{equation}%
The parameter $\zeta $ can be found from the condition for the extremum of
the action Eq. (\ref{43}),
\begin{equation}
\zeta =\frac{\sqrt{p_{\phi _{k}}^{2}}}{2\sqrt{3}\pi a\sqrt{\lambda }},
\label{49}
\end{equation}
and $\lambda $ can be found from Eq. (\ref{46}) as a function (functional) of $\phi
_{k}\left( s\right) $. However, if we continue to use the quasiclassical
approximation to calculate the path integral Eq. (\ref{47}), the arbitrary
trajectory of the scalar fields $\phi _{k}\left( s\right) $ in Eq. (\ref{46})
should be replaced by the classical trajectory of their motion (together
with $a$ and $P_{\varepsilon }$). In the theory of inflation \cite{16}, it
is customary to consider this dynamics, starting from the stage when the
potential energy $V\left( \phi \right) $ is maximum. However, we will move
the time reference closer to the singularity, assuming that inflation was
preceded by a stage of rapid change of fields, so,
\begin{equation}
\overset{\cdot }{\phi }_{k}^{2}>>V\left( \phi \right) .  \label{50}
\end{equation}%
In this case,
\begin{equation}
\zeta \cong \frac{\sqrt{p_{\phi _{k}}^{2}}}{2\pi \sqrt{\overset{\cdot }{\phi
}_{k}^{2}}},  \label{51}
\end{equation}%
and the Schr\~{o}dinger equation, corresponding in the quantum communication
theory Eq. (\ref{48}), takes the form:
\begin{equation}
i\hbar \frac{\partial \psi }{\partial \sigma }=\frac{\hbar }{2\pi }\sqrt{%
-\Delta _{\phi _{k}}}\psi ,  \label{52}
\end{equation}%
where under the root sign is the Laplace operator in the configuration space
of scalar fields, and
\begin{equation}
d\sigma =\sqrt{d\phi _{k}^{2}}  \label{53}
\end{equation}%
is an element of length in this space. Thus, the initial state of the
universe near a singularity is described by a de Broglie plane wave in the
configuration space of matter fields with initial momenta $p_{\phi _{k}}$
and energy
\begin{equation}
W=\left\vert \left\vert p_{\phi }\right\vert \right\vert =\sqrt{p_{\phi
_{k}}^{2}}.  \label{54}
\end{equation}%
To find the energy of the space $P_{\varepsilon }$ near the singularity, we
simplify the consideration further by restricting ourselves to a single
scalar field $\phi $. From the equation of motion of this field (taking into
account Eq. (\ref{50})),
\begin{equation}
\overset{\cdot \cdot }{\phi }+3\frac{\overset{\cdot }{a}}{a}\overset{\cdot }{%
\phi }=0 , \label{55}
\end{equation}%
we find
\begin{equation}
\overset{\cdot }{\phi }=\frac{\Phi }{a^{3}}.  \label{56}
\end{equation}%
After this, from Eq. (\ref{36}) we obtain
\begin{equation}
P_{\varepsilon }=\frac{\Phi ^{2}}{a^{3}}.  \label{57}
\end{equation}%
Thus, at the beginning $a=0$, we have a singularity: the energy of space
there is infinite. It with expansion decreases according to Eq. (\ref{57}). The
energy of space reaches its minimum value at the boundary of the inflation
stage when $(\overset{\cdot }{\phi })^{2}=V\left( \phi \right) $. What follows
is an exponential increase in the energy of space at the inflation stage.
Starting from the indicated minimum value, the energy of the space $%
P_{\varepsilon }$ can be considered as a physical parameter of time \cite{12}.

\section{CONCLUSIONS}

The variant of covariant quantum theory proposed in this paper and its
subsequent modification with additional conditions on the dynamics of
time-like degrees of freedom allow us to solve the problem of time in it. On
the one hand, proper time is clearly added to the dynamic variables of the
theory at the classical level. On the other hand, in quantum theory, this
time disappears as a result of FS integration (the condition for the
propagator covariance). Its dynamic meaning is restored by the conditions of
the classical dynamics of time-like degrees of freedom. These conditions
lead to the appearance of an additional dynamic variable - self-energy, for
which the classical law of its change at the classical and quantum level is
formulated. Thus, the modification is focused here on time-like degrees of
freedom of the system (coordinate $x_{0}$ of Minkowski space and radius $a$
in the mini-superspace of the Friedmann universe) and represents a variant
of combining the principles of covariance and quantization.

\section{Acknowledgements}
The authors would like to thank A. V. Goltsev and V. A. Franke for useful discussions.

%%%\noindent $^{\ast }$ E-mail address: alex.lukyan@rambler.ru

%%%\noindent $^{+}$ E-mail address: inna.lukyan@mail.ru

%\begin{references}


\begin{thebibliography}{99}
\bibitem{1} E. S. Fradkin and G. A. Vilkovisky, \emph{Phys. Lett}. \textbf{55B}, 224 (1975).

\bibitem{2} I. A. Batalin and G. A. Vilkovisky, \emph{Phys. Lett}. \textbf{69B}, 309 (1977).

\bibitem{3} M. Henneaux, \emph{Physics Reports} \textbf{126}, 1 (1985).

\bibitem{4} T. Kugo and I. Ojima, \emph{Suppl. Progr. Theor. Phys}. \textbf{66}, 1 (1979).

\bibitem{5} F. R. Ore and P. van Nieuwenhuisen, \emph{Nucl. Phys.} B \textbf{204}, 317 (1982).

\bibitem{6} Jan Govaerts, A note of the Fradkin-Vilkovisky theorem, CERN-TH
5010/88 (1988).

\bibitem{7} V. A. Fock, \emph{Izv. Acad. Nauk SSSR}, 551 (1937).

\bibitem{8} J. Schwinger, On gauge invariance and vacuum polarization, \emph{Phys.
Rev.} \textbf{82}, No 5, 664--679 (1951).

\bibitem{9} B.S. De Witt, in General Reletivity, An Einstein sentenary survey
edited by S.W. Hawking and W. Israel, Cambridge University Press, Cambridge,
463 (1983).

\bibitem{10} J. D. Bjorken, Sidney D.Drell, Relativistic quantum fields,
Mc Graw-Hill Book Company (1978).

\bibitem{11} C. W. Misner, K. S. Thorne and J. A. Wheeler,
Gravitation, W. H. Freeman and Company, San Francisco (1973).

\bibitem{12} N. N. Gorobey and A. S. Lukyanenko,  St. Petersburg Politechnical 
State University Journal \emph{Physics and Mathematics} \textbf{11}, No 1, 147-156 (2018).
%\emph{Int. Journ of Modern Physics} A \textbf{31}, No 2\&3, 1641014 (2016).

\bibitem{13} L. D. Faddeev and A. A. Slavnov, Gauge fields: An introduction to
quantum theory, Westview Press, 2d edition, 236 (1993).

\bibitem{14} R. Arnovitt, S. Deser and C. Misner, Dinamical Structure and
Definition of Energy in General Relativity, \emph{Phys. Rev}. \textbf{116}, No 5,
1322-1330 (1959).

\bibitem{15} J. B. Hartle, S. W. Hawking and T. Hertog,
arXiv:0803.1663v1 [hep-th] 11 Mar 2008.

\bibitem{16} A. Linde, Inflation, Quantum Cosmology and the Anthropic
Principle, In \textquotedblleft Science and Ultimate Reality: From Quantum
to Cosmos\textquotedblright , honoring John Wheeler's 90th birthday. J. D.
Barrow, P. C. W. Davies, \& C. L. Harper eds. Cambridge University Press (2003).
\end{thebibliography}
\end{document}